\documentclass[aps,prl,twocolumn,showpacs,floatfix]{revtex4-1}

\usepackage{amssymb,amsmath,epsfig,graphicx}
\begin{document}

\author{L.~Dinis$^{1,2,3}$, P.~Martin$^1$, J.~Barral$^1$, J.~Prost$^{1,4}$,  J.F.~Joanny$^1$}
\affiliation{$^1$Laboratoire Physico-Chimie Curie, CNRS, Institut Curie, UPMC, 26 rue d'Ulm, F-75248 Paris Cedex 05, France}
\affiliation{$^2$Grupo Interdisciplinar de Sistemas Complejos (GISC)}
\affiliation{$^3$Universidad Complutense de Madrid, Spain}
\affiliation{$^4$E.S.P.C.I., 10 rue Vauquelin, 75231 Paris Cedex 05, France}

\title{Fluctuation-response theorem for the active noisy oscillator of the hair-cell bundle}
\begin{abstract}
The hair bundle of sensory cells in the vertebrate ear provides an example of  a noisy
oscillator close to a Hopf bifurcation. The analysis of the data from both spontaneous
and forced oscillations shows a strong violation of the fluctuation-dissipation theorem,
revealing the presence of an underlying active process that keeps the system out of
equilibrium. Nevertheless, we show that a generalized fluctuation-dissipation theorem,
valid  for non-equilibrium steady states, is fulfilled within the limits of our experimental accuracy and computational
approximations, when the adequate
conjugate degrees of freedom are chosen.

\end{abstract}
\maketitle

The fluctuation-dissipation theorem (FDT) is the cornerstone of linear response theory
for systems at thermal equilibrium \cite{Callen1951}:
it relates the response to small perturbations  to the correlations
of spontaneous fluctuations and
connects the microscopic dynamics of the system to the
macroscopic transport coefficients, such as  diffusion constant, conductivity,
absorption rates, etc.

Many systems operate
far from thermodynamic equilibrium and therefore do not obey
the fluctuation-dissipation theorem. One illustrative example is given by the hair-cell bundle. The hair bundle operates as a mechanical antenna that protrudes from the apical surface of each hair cell in the
inner ear of vertebrates \cite{hudspeth_89,barral_physical_2011}. Hearing starts when sound-evoked deflections of this organelle are transduced into electrical signals that then travel to the brain. Composed of cylindrical protrusions - the stereocilia - that are arranged in rows of increasing heights, the hair bundle displays a staircase pattern. Stereocilia are interlinked near their tips by fine oblique filaments. Tip-link tension controls the open probability of mechanosensitive ion channels. The hair cell can power noisy spontaneous oscillations of its hair bundle that display a
spectacular violation of the FDT \cite{Martin2001}. The behavior of the hair bundle has been described by the
generic normal form of a dynamical system that operates on the stable side of a Hopf
bifurcation \cite{Juelicher2009}. In this letter, we focus on this particular class of out-of-equilibrium systems.

Several generalizations of the FDT to non-equilibrium systems have been proposed \cite{Agarwal1972,Chetrite2008,Speck2006,Baiesi2009}. The generalized fluctuation-dissipation theorem (GFDT) of Prost {\em et al.~}\cite{Prost2009} applies to systems with Markovian
dynamics in a non-equilibrium steady state. Applying the GFDT to experimental
measurements on the hair bundle, we show here that a proper choice of variables restores a
relation between spontaneous fluctuations and linear response.

\begin{figure}[h]
\centering
\includegraphics[width=\linewidth]{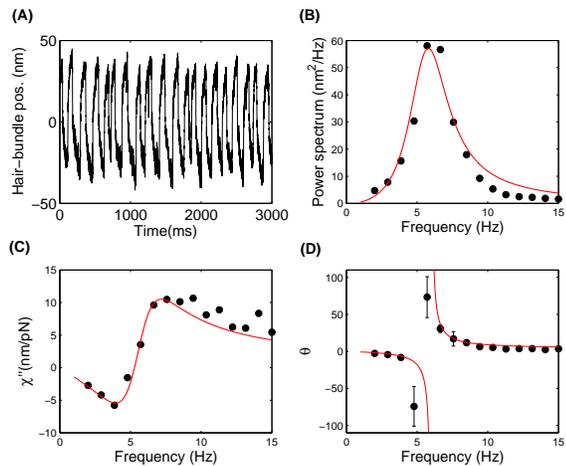}
\caption{
\label{fig_global_fit} Experimental data on a hair bundle: (A) Spontaneous hair-bundle oscillation.
(B) Power spectral density $\tilde C_{xx}(\nu)$ averaged from 15 different trajectories as a function of frequency $\nu$.
(C) Imaginary part of the response function $\tilde\chi''_{xx}(\nu)$.
Thin red lines (B-C) correspond to a
simultaneous fit of $\tilde C_{xx}(\nu)$ and  $\tilde \chi''_{xx}(\nu)$
to theoretical expressions derived from Eq. \eqref{eq_xy}.
(D) Fluctuation-response ratio
$\theta$ as defined in
Eq. \eqref{theta} fitted to theoretical expression given in Refs. \cite{Martin2001}
}
\end{figure}

Details of the experiment are found in Refs.~\cite{Martin1999,Martin2001,Tinevez2007}.
The oscillatory movement of a hair bundle was monitored with a glass fiber attached to
its tip (Fig.~\ref{fig_global_fit}A). The
fiber was used both to exert sinusoidal forces and to report hair-bundle noisy oscillations. The power spectrum  $\tilde C_{xx}(\omega)=\int C_{xx}(t) \ e^{i\omega t} dt$ of
spontaneous hair-bundle position $x$, which is the Fourier transform
of the correlation function $C_{xx}(t)=\langle x(t)x(0)\rangle$, displays a broad peak
centered at a characteristic
frequency $\nu_0=\omega_0/2\pi\simeq 6\text{Hz}$ (Fig.~\ref{fig_global_fit}B).
For stimulation by external
sinusoidal forces $f(t)$, the linear response function $\tilde \chi = \tilde\chi' + {\rm i}\tilde\chi''$ is defined at each angular frequency $\omega$ by: $\langle \tilde
x(\omega)\rangle=\tilde\chi(\omega)\tilde f(\omega)$, where tildes denote Fourier
components.  Its imaginary part $\tilde\chi''(\omega)$ is
proportional to the work received by the system from the external force for stimulation at a frequency $\omega$ \cite{Chandler}. At thermal equilibrium, with our definition of the Fourier transform, $\tilde\chi''(\omega)$
must always be positive (for $\omega \geq 0$). Remarkably, in the
case of the oscillatory bundle, $\tilde\chi''(\omega)$ changes sign near $\nu_0$, as shown in Fig.~\ref{fig_global_fit}C. At low frequencies, the work received by the bundle is negative, meaning that energy is transferred from the hair bundle to the fiber.
An energy consuming or active process must thus be at work to power hair-bundle movements.

At thermal equilibrium, the FDT relates
the imaginary part of the response function to the power spectrum of spontaneous fluctuations for a degree
of freedom $x$
\begin{equation}
\tilde C_{xx}(\omega)=2kT\frac{\tilde \chi''_{xx}(\omega)}{\omega},
\label{eq_FDT}
\end{equation}
 where $T$ is the temperature
 and $k$ the Boltzmann constant. Departure from equilibrium can be characterized by the fluctuation-response ratio
 \begin{equation}
 \label{theta}
 \theta=\frac{\omega\tilde C_{xx}(\omega)}
{2kT\tilde \chi_{xx}''(\omega)},
\end{equation}
sometimes called the effective temperature (in units of the actual
temperature T).
This ratio equals one when the system
is at equilibrium. In an out-of-equilibrium system, $\theta$
might depend on frequency and be either positive or negative.
For the hair bundle, the fluctuation-response ratio $\theta$ shown in
Fig.~\ref{fig_global_fit}D depends on frequency and presents a striking divergence
in the vicinity of $\nu_0=\omega_0/2\pi$, corresponding to the sign change of
$\tilde \chi''_{xx}(\omega)$. However, if the GFDT applies, a fluctuation-response relation is restored with an appropriate choice of the conjugate variable $X$ to the external force \footnote{Note that we used here a different sign convention than in \cite{Prost2009}, both for the definition of the Fourier transform and the definition of the variable $X$.}:
\begin{equation}
\label{eq_GFDT}
\tilde\chi_{XX}(\omega)-\tilde\chi_{XX}(-\omega) = {\rm i}\omega \tilde C_{XX}(\omega) .
\end{equation}
The behavior of the hair bundle for small deflections has been described as a
two-variable dynamical system:
\begin{equation}
\frac{\rm d}{{\rm d}t}\left(
\begin{array}{c}
x\\
y
\end{array}
\right)=\left(
\begin{array}{cc}
-r & \omega_0\\
-\omega_0 &-r
\end{array}
\right)\left(
\begin{array}{c}
x\\
y
\end{array}
\right)+\left(
\begin{array}{c}
f_x\\
0
\end{array}
\right)
+\left(
\begin{array}{c}
\eta_x\\
\eta_y
\end{array}
\right).
\label{eq_xy}
\end{equation}

The variable $x$ is the deflection of the hair bundle, $r=k/\lambda$ is a damping rate
where $\lambda$ and $k$ are respectively the effective drag coefficient and the stiffness
of the bundle, $f_x=f_\text{ext}/\lambda$ where $f_\text{ext}$ is  the external force on the
hair bundle. The second degree of freedom  $y$ is related
to the force exerted by the active process within the hair bundle and couples
to the displacement $x$. The noises $\eta_x$ and $\eta_y$ describe fluctuations in the system.
We treat the two Langevin forces
as white noises so that the dynamical system is Markovian. At the low frequencies of the experiment ($\sim 10$ Hz), we expect noise correlation times to be sufficiently short that the noises can be considered as delta-correlated. Non-Markovian effects are expected at higher frequencies only, as discussed below. Equation  \eqref{eq_xy} is to be understood as a renormalized expression, valid for providing two point correlation functions and linear responses, of a more complex non-linear problem ~\cite{Nadrowski2004,Tinevez2007,Juelicher2009}.
  As a result, the noises $\eta_x$ and $\eta_y$ are in general correlated. However, experimentally, the cross-correlation turned out to be very small and the two noises are effectively independent.
The noise correlations are written as $\langle \eta_x(t)\eta_x(t')\rangle=\sigma_{\eta_x}
\delta(t-t')$, $\langle \eta_y(t)\eta_y(t')\rangle=\sigma_{\eta_y}\delta(t-t')$.

The dynamical
system described by Eq.\eqref{eq_xy} is the canonical form of a system close
to a Hopf bifurcation \cite{Strogatz1997}. If $r>0$, it displays damped
spontaneous oscillations of
frequency $\omega_0$.
The expressions for the power spectrum and the complex response function to an external
force $f_x$ can be readily computed from this model and were used for a global fit
of the experimental data
with a unique set of parameters $r$, $\omega_0$, $\sigma_{\eta_x}$ and $\sigma_{\eta_y}$
in Fig.~\ref{fig_global_fit}
(the real part of the response function is not shown).

With the choice of $x$ as conjugate variable of the external force $f_x$,
the fluctuation-dissipation theorem is violated (Fig.\ref{fig_global_fit}C). This is a strong signature of a
non-equilibrium behavior. Nevertheless, the dynamics of \eqref{eq_xy} being Markovian,
the generalized fluctuation-dissipation theorem (GFDT) of Prost {\em et
al.}~\cite{Prost2009} holds for the correct conjugate variable $X$
of the external force. In the case of the two-dimensional linear system at hand, Eq. 5 in Ref. ~\cite{Prost2009} yields:
\begin{eqnarray}
\label{eq_XY}
\left(
\begin{array}{c}
X\\
Y
\end{array}
\right)
=[A^{-1}]^{\rm T}\Sigma_A^{-1}\left(
\begin{array}{c}
x\\
y
\end{array}
\right)
\end{eqnarray}
with
\begin{equation}
A=-\left(
\begin{array}{cc}
-r & \omega_0\\
-\omega_0 & -r
\end{array}
\right),
\Sigma_A=\left(
\begin{array}{cc}
\langle x^2 \rangle_\text{ss} & \langle x y\rangle_\text{ss}\\
\langle x y\rangle_\text{ss} & \langle y^2\rangle_\text{ss}
\end{array}
\right),
\end{equation}
where the averages in the matrix $\Sigma_A$ are calculated in the steady state.
A direct test of the GFDT would thus require a measurement of the
internal degree of freedom $y$, which is not experimentally accessible.

To circumvent this limitation, we propose three different approaches. On the one hand,
using the measured $x$ value, we
estimate the hidden variable either by computing the linear combination $z=y\omega_0-rx$ of $x$ and
$y$ using a denoising procedure, or by an optimization technique.
On the other hand, we directly evaluate the correlations
involving $z$ which are sufficient to test the validity of the GFDT.
Using the variable $z$, we write the dynamical system as
\begin{eqnarray}
\frac{\rm d}{{\rm d}t}\left(
\begin{array}{c}
x\\
z
\end{array}
\right) & = &\left(
\begin{array}{cc}
0 & 1\\
-(r^2+\omega_0^2) & -2r
\end{array}
\right)\left(
\begin{array}{c}
x\\
z
\end{array}
\right)
+\left(
\begin{array}{c}
f_x+\eta_x\\
f_z+\eta_z
\end{array}
\right)\nonumber \\
& \equiv & -R
\left(
\begin{array}{c}
x\\
z
\end{array}
\right)
+\left(
\begin{array}{c}
f_x+\eta_x\\
f_z+\eta_z
\end{array}
\right),
\label{eq_system_z}
\end{eqnarray}
where the noise and force in the $z$ equation are  $\eta_z=-r\eta_x+\omega_0 \eta_y$
and $f_z=-rf_x$.

In the absence of external force, $\frac{{\rm d}x}{{\rm d}t}=z+\eta_x$.
We can therefore estimate the value of $z$ by filtering the time series of the speed data,
eliminating the high frequency noise: at each point of the trajectory, the value of the speed is calculated by averaging over the $N$ preceding points, where  $N$ is  such that the averaging effectively filters signals faster than 60Hz.
This frequency is several times higher than the spontaneous oscillation frequency of the bundle $\nu_0\simeq6\text{Hz}$ and could be varied without much effect on the final results as long as it is high enough ($\geq 30$ Hz) to preserve the waveform of hair-bundle oscillation and low enough ($\leq 90$ Hz) to get rid of most of the high-frequency noise. Denoising implicitly relies on the assumption that the velocity $\dot x$ can be split into a
variable $z$ with exponentially decaying  correlations plus white noise. As can be checked
at very low frequencies, it only
gives an approximation $z_e$ of the actual variable $z$.

Once the variables $x$ and $z_e$ are obtained, we apply the GFDT to
the system described by equation \eqref{eq_system_z}, which is also Markovian.
The correlation matrix for the $x$ and $z$  variables in Fourier
space is then approximated by
\begin{equation}
\tilde C_{\vec x \vec x}(\omega)\simeq
\left(
\begin{array}{cc}
\langle \tilde x(\omega)\tilde x^*(\omega)\rangle &\langle \tilde x(\omega)\tilde z_e^*(\omega)\rangle \\
\langle \tilde z_e\omega)\tilde x^*(\omega)\rangle &\langle \tilde z_e(\omega)\tilde z_e^*(\omega)\rangle
\end{array}
\right),
\end{equation}
where the star denotes a complex conjugate.

We compute $\tilde x(\omega)$
and $\tilde z_e(\omega)$ using the FFT algorithm on the experimental data.
The matrix $R$ is obtained from the values of $r$ and $\omega_0$  and then used
to perform the change of variables
\begin{equation}
\vec X\equiv\left(
\begin{array}{c}
X\\
Z_e
\end{array}
\right)=[R^{-1}]^{\rm T}\Sigma^{-1}
\left(
\begin{array}{c}
x\\
z_e
\end{array}
\right)
\end{equation}
where $\Sigma=
\left(
\begin{array}{cc}
\langle x^2 \rangle_\text{ss} & \langle x z_e\rangle_\text{ss}\\
\langle z_e x\rangle_\text{ss} & \langle z_e^2\rangle_\text{ss}
\end{array}
\right)$.
In the new variables, the power spectrum reads
\begin{equation}
\tilde C_{\vec X\vec X}(\omega)=[R^{-1}]^{\rm T}\Sigma^{-1}\tilde C_{\vec x\vec x}(\omega)[\Sigma^{-1}]^{\rm T}R^{-1}
\label{eq_Cxpxp}
\end{equation}
and the response function
\begin{equation}
\tilde \chi_{\vec X\vec X}(\omega)=[R^{-1}]^{\rm T}\Sigma^{-1}
\tilde\chi_{\vec x\vec x}(\omega)=[R^{-1}]^{\rm T}\Sigma^{-1}[R+{\rm i}\omega]^{-1}.
\end{equation}

The GFDT \cite{Prost2009} then imposes a relation between fluctuations and response given by Eq.\ref{eq_GFDT}.
In particular, for the first diagonal element, we expect the fluctuation-response ratio:
\begin{equation}
\theta= \frac{\omega\tilde C_{XX}(\omega)}
{2\tilde\chi''_{XX}(\omega) }=1.
\label{eq_Teff}
\end{equation}

\begin{figure}
\centering
\includegraphics[width=\linewidth]{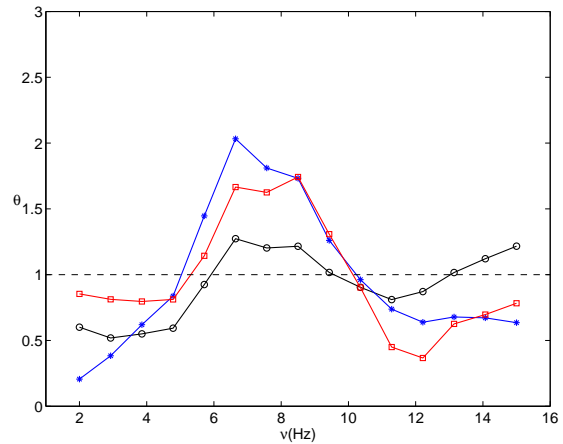}
\caption{ \label{fig_Teff} The fluctuation-response ratio $\theta$ vs.~stimulation frequency $\nu$ using the three
different methods explained in the text.
(black) $\circ$:  denoising of $z$, (red) $\square$: estimation of $\tilde C_{\vec x\vec z}
(\omega)$, (blue) *: $y$ estimation by maximization of probability. Note that
the power spectra were smoothed out to eliminate
some of the noise by a moving average algorithm. Lines are just guides for the eye.}
\end{figure}

In Fig.~\ref{fig_Teff} (black circles), we plot the fluctuation-response ratio $\theta$ evaluated from the experimental data.
We find that $\theta$ wiggles about the value $1$, within a range that stretches from $0.5$ to $2$. This is a remarkable behavior considering that, when fluctuations and response were evaluated with the hair-bundle position $x$ as the relevant degree of freedom, the fluctuation-response ratio changed sign and diverged near the characteristic frequency of spontaneous oscillations (Fig.~\ref{fig_global_fit}D). Although the GFDT imposes that $\theta$ be precisely equal to $1$, numerical simulations shown below demonstrate that the experimental data are compatible with the GFDT.

We then used an inference method to estimate
the variable $y$
from the measured trajectories. The assumption of Gaussian white noises for $\eta_x$ and $\eta_y$ in
Eq.\eqref{eq_xy} implies that the combinations $\dot x+rx-\omega_0y$ and $\dot
y+\omega_0y+rx$ are Gaussian variables for spontaneous oscillations ($f_x=0$).
Discretizing the evolution equation \eqref{eq_xy} in $N$ time steps $\Delta t$, we
find
\begin{eqnarray}
x_{n+1}-x_n+\Delta t(rx_n-\omega_0y_n)&\sim \mathcal{N}(0,{\sigma_{\eta_x}
\Delta t})\nonumber \\ y_{n+1}-y_n+\Delta t(rx_n-\omega_0y_n)&\sim
\mathcal{N}(0,{\sigma_{\eta_x}\Delta t})
\end{eqnarray}
where $\mathcal N(\mu,\sigma^2)$ denotes the normal distribution of average $\mu$ and
variance $\sigma^2$.
The probability $\rho(\{x_n,y_n\})$ of observing a discrete full trajectory $\{x_n,y_n\}_{n=1}^{n=N}$ is then a
product of $2N$ Gaussian distributions. Maximizing this probability with respect to
the
$y_n$
variables ($\partial \rho/\partial y_n=0,\forall n$) gives a linear system of equations for the most likely
trajectory $\{y_n\}$ in terms of the measured variable
$\{x_n\}$ and the parameters $r,\omega_0, \sigma_{\eta_x}, \sigma_{\eta_y}$. We use
this estimated trajectory to perform the change of variables (Eq. \eqref{eq_XY}) required for the GFDT. The resulting  $\theta$ is depicted  in Fig. \ref{fig_Teff}.

Our third approach to test the GFDT avoids any $y$ estimation by directly
calculating the correlation matrix from the measured data.
Only the first element of the matrix $\tilde C_{xx}(\omega)$ can be directly
obtained from the experimental data. To estimate the elements involving $z$, we proceed
as follows. Fourier transforming \eqref{eq_system_z} for $f_x=0$ we get $\tilde z(\omega)=
{-\rm i}\omega\tilde x(\omega)-\tilde \eta_x(\omega)$, which we use to calculate
the cross-correlation
\begin{equation}
\tilde C_{xz}(\omega)=
{\rm i}\omega \tilde C_{xx}(\omega)-\langle \tilde x(\omega)\tilde \eta_x(-\omega)\rangle.
\label{eq_cxz}
\end{equation}
The second term in \eqref{eq_cxz} is evaluated by  means of the evolution equation \eqref{eq_system_z} giving
\begin{equation}
\tilde C_{xz}(\omega)={\rm i}\omega \tilde C_{xx}(\omega)-\sigma_{\eta_x}\frac{r-{\rm i}\omega}{r^2+\omega_0^2-\omega^2-2r{\rm i}\omega}
\end{equation}
where the only unknown parameter is the noise intensity $\sigma_{\eta_x}$. However,
from \eqref{eq_system_z} one can show that
$\sigma_{\eta_x}=-2\Sigma_{12}$
Following the same lines both $\tilde C_{zx}(\omega)$ and $\tilde C_{zz}(\omega)$
are expressed in terms of $\tilde C_{xx}(\omega)$, $r, \omega_0$ and $\Sigma_{12}$.
Finally, we estimate $\Sigma$ from the data by noting that
$\Sigma_{11}=\langle x(0)^2\rangle=C_{xx}(t=0)$, $
\Sigma_{12}=\langle x(0)z(0)\rangle=\left.\frac{{\rm d}C_{xx}(t)}{d{\rm t}}
\right|_{t=0}\label{eq_dcdt}$ or alternatively by fitting the power spectrum expressed as a
function of $\Sigma_{11}$ and $\Sigma_{12}$, noting that
$\Sigma_{22}=(r^2+\omega_0^2)\Sigma_{11}$.

Once $\Sigma$, $R$ and $\tilde C_{\vec x\vec x}(\omega)$ are known we insert them
into equation \eqref{eq_Cxpxp}
and compute the fluctuation-response ratio $\theta$
as in Eq.\eqref{eq_Teff} (Fig.~\ref{fig_Teff}).

\begin{figure}
\centering
\includegraphics[width=\linewidth]{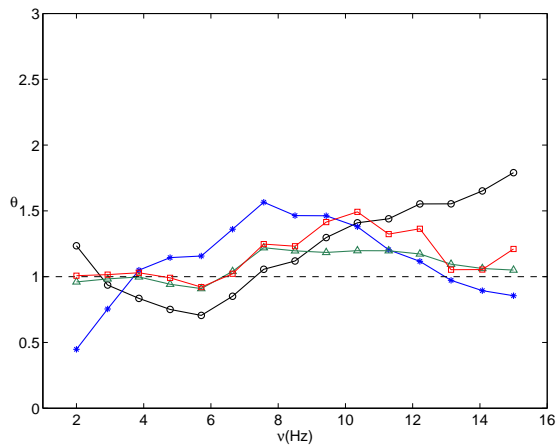}
\caption{\label{fig_sim}$\theta$ ratio computed after the change of variables prescribed by the GFDT, from simulated data. Computed using 
(black) $\circ$: $\tilde x$ and $\tilde z_e$, (blue) $*$: $\tilde x$ and $\tilde y$ estimated by probability maximization, (red) $\square$: direct $\tilde C_{\vec x\vec y}(\omega)$ estimation and (green) $\triangle$: $\tilde x$ and
actual (no filtering nor estimation) $\tilde y$. Simulations where done with $r=10.2\text{s}^{-1}$, $\omega_0=36.5\text{rad/s}$ and $\sigma_x=10000\text{nm}^2/s$ and $\sigma_y=10000\text{nm}^2/s$. Lines are just guides for the eye.}
\end{figure}

In order to asses the impact of the three different
estimation methods, we performed numerical simulations with parameters similar to those
of the experiment and repeated the procedure using both our estimates
and the actual $y$ value, which is available in simulations.
The simulations were performed using the Euler-Mayurama method \cite{Manella2000}
to solve equation \eqref{eq_xy}.
As expected, results in Fig.~\ref{fig_sim} show that the agreement with the theorem is best
when the actual variable $y$ is used. However, even then, we still observe deviations of $\theta$ by about 25\% due to a lack of averaging.
In addition, both the moving-average procedure and the inference method imply a dependence
on past history, and thus introduce some degree of non-Markovianity that might explain
further departure from the GFDT. Because experiments and simulations show similar deviations of the fluctuation-response ratio from $1$, we consider that it is as close to 1 as possible, in view of the inherent limitations associated with the methods that we used to estimate this ratio.

In conclusion, we showed that the generalized fluctuation-dissipation theorem \cite{Prost2009} applies to oscillatory hair-cell bundles. Although the hair bundle provides a compelling example of a complex biological system that operates away from thermal equilibrium, its linear mechanical response is related to
steady-state fluctuations with the appropriate choice of a conjugate variable to the external force. This relation holds for frequencies close to the frequency of spontaneous oscillation, at which the hair bundle can be described by a two-dimensional dynamical system operating near a Hopf bifurcation. This property affords a means to estimate the hidden variable that underlies the activity of the hair bundle.
Because the hair bundle must satisfy the hypotheses of the GFDT, our results support the description of the hair bundle as a single noisy
oscillator governed by Markovian dynamics and therefore go against a viscoelasticity of the
hair bundle in the range of frequencies that we studied.
At higher frequencies, however, the hair bundle could  become non-Markovian, due mainly to memory  resulting from visco-elasticity \cite{Kozlov2012}
 or from colored fluctuations in the opening and closing of the transduction channels \cite{Nadrowski2004}.
Channel clatter is only expected at frequencies above $\sim
1$ kHz
\cite{Nadrowski2004}, where a departure from the GFDT could be observed.

Our work relates to the experiments of Ref. \cite{Gomez-Solano2011} which test
the same generalized fluctuation-dissipation theorem for an experimental system consisting of a Brownian particle in a toroidal optical trap. In contrast to our study where we have to assume a Hopf bifurcation dynamics with noise, in the optical trap experiment the evolution equation is known, as the potential felt by the particle is also applied using the trap.

We have provided  three methods for the estimation of correlations involving
the non-measured degree of freedom. Both the denoising and the inference methods can be
directly applied to other noisy systems. It would be however desirable to
perform experiments where, in addition to the displacement, the dynamics of the active term
can be controlled and measured. A good candidate for an additional measurement is
the ionic current that flows through the bundle, which is known to influence either the
myosin motors that generate the force inside the bundle or the transduction channel to
which the motors are attached \cite{Duke2003}.

We would like to thank F. J\"ulicher for discussions.
L.D.~acknowledges support from grants MOSAICO and ENFASIS (FIS2011-22644) and the POSDEXT-MEC program (Spanish Government) and Universidad de Sevilla.

\end{document}